# Dielectric catastrophe at the magnetic field induced insulator to metal transition in $Pr_{1-x}Ca_xMnO_3$(x = 0.30, 0.37) crystals

B. Kundys[a], N. Bellido, C. Martin, and Ch. Simon

Laboratoire CRISMAT, CNRS UMR 6508, ENSICAEN, 6 Bd. du Maréchal Juin, 14050 Caen Cedex, France

**Abstract.** The dielectric permittivity and resistivity have been measured simultaneously as a function of magnetic field in $Pr_{1-x}Ca_xMnO_3$ crystals with different doping. A huge increase of dielectric permittivity was detected near percolation threshold. The dielectric and conductive properties are found to be mutually correlated throughout insulator to metal transition evidencing the dielectric catastrophe phenomenon. Data are analyzed in a framework of Maxwell-Garnett theory and the Mott-Hubbard theory attributed to the role of strong Coulomb interactions.

## 1 Introduction

The dielectric properties of materials are of great importance from both fundamental and application points of view. Materials in which metal to insulator (IM) transition (or vice versa) takes place are of particular interest for basic investigations of contemporary condensed matter physics. The dielectric permittivity is the parameter characterizing sample's response to an applied electric field. Near the percolation threshold dielectric permittivity increases throughout IM transition if the transition is approached from the insulating side [1]. This phenomenon was called "dielectric catastrophe" by Mott [1] and was quite rarely evidenced in liquids [1,2], bulks [3,4] and thin films [5,6]. The theoretical aspect of dielectric and conductive properties of the system, in which conducting phases are embedded into insulating matrix has also received some interest [7–10]. It was also logically predicted that dielectric permittivity tends to diverge in any case at a percolation threshold of IM transition and does not depend on its direction [11]. Here a distinctive evidence for such prediction is presented in $Pr_{1-x}Ca_xMnO_3$ (x = 0.30, 0.37) compounds, which are an excellent material for this study as they reveal magnetic field [12], pressure [13], light [14], current [15] and X-ray [16] — induced unconventional IM transition. It also has to be noted that (controversial) anomalies of dielectric permittivity as a function of temperature were previously observed $Pr_{0.67}Ca_{0.33}MnO_3$ crystal [17,18].

In this work, we have measured, dielectric permittivity simultaneously with resistivity while IM transition was induced by external magnetic field.

## 2 Experimental details

Single crystals having several-cm-long size were grown by the floating-zone method in an image furnace using feeding rods of the nominal compositions of $Pr_{0.7}Ca_{0.3}MnO_3$ and $Pr_{0.6}Ca_{0.4}MnO_3$. The samples were appropriately cut from the middle part of those crystals. The electron diffraction showed the existence of twinning structure in both specimens. Therefore, physical measurements performed on our samples should be averaged over the six oriented domains which coexist in the *Pnma* phase. The Ca doping concentration was checked by dispersive spectroscopy analysis, and the cationic composition was found to be x = 0.30 and 0.37, respectively. Experiments were performed in the PPMS Quantum Design cryostat. The capacitance and resistance were measured concurrently using Agilent 4248A RLC auto balance bridge at 1 MHz. The magnetic field was applied perpendicular to the direction of the electric field. Indium was used to make electrical contacts. Samples were systematically cooled from 300 K in zero magnetic field before measurements were done.

## 3 Results and discussions

In a zero magnetic field both samples are good insulators at low temperature. As can be seen from Figures 1

[a] e-mail: Bohdan.Kundys@ensicaen.fr

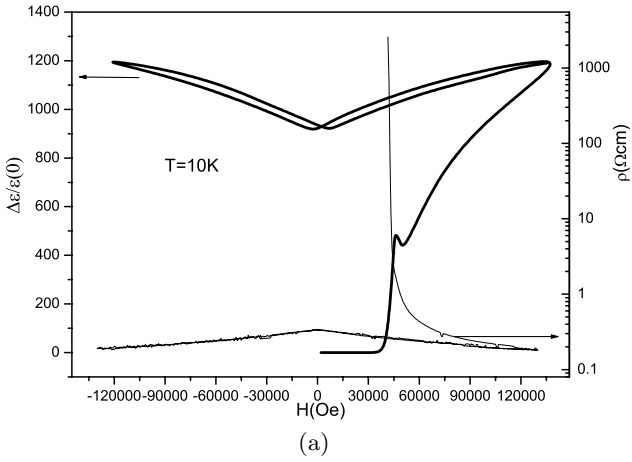

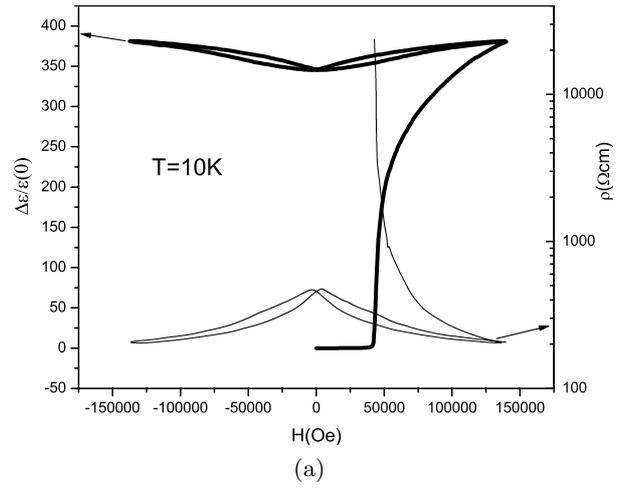

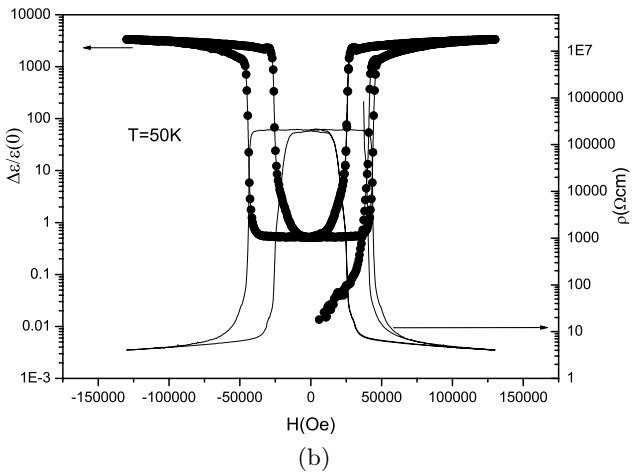

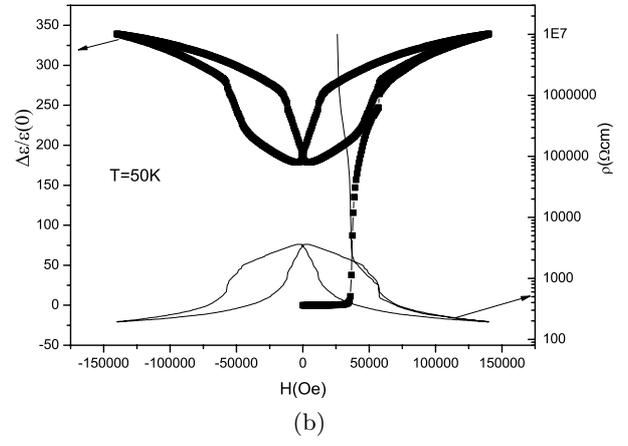

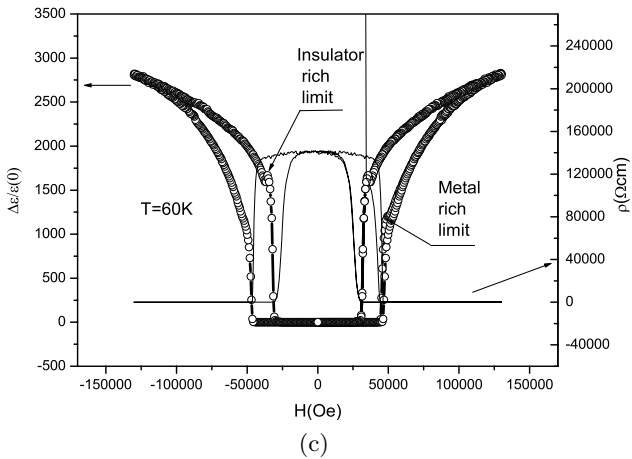

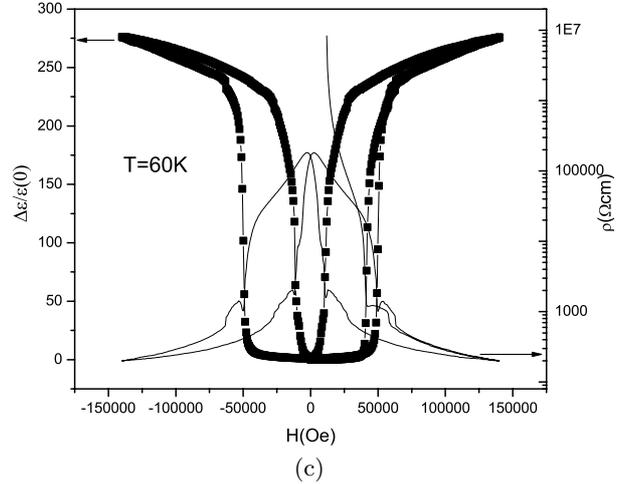

**Fig. 1.** Magnetodielectric and resistivity isotherms $Pr_{0.63}Ca_{0.37}MnO_3$ crystal taken at 10, 50 and 60 K.

**Fig. 2.** Resistivity and magnetodielectric loops measured for $Pr_{0.7}Ca_{0.3}MnO_3$ crystal at 10, 50 and 60 K.

and 2, the magnetic field dramatically increases the dielectric permittivity through IM transition and decreases it for MI transition in both compounds. The resistivity behaves in an opposite way. For instance, at 10 K dielectric permittivity rises from 35 up to approx. 42 000 at 14 T for $Pr_{0.63}Ca_{0.37}MnO_3$ compound and from 14 to approx. 5400 in $Pr_{0.7}Ca_{0.3}MnO_3$ sample (Figs. 1a, 2a). Quite similar, albeit much smaller resistivity correlated dielectric behaviour was also observed in $Nd_{1/2}Sr_{1/2}MnO_3$ compound using magneto-optical studies [19]. Using effective media approximation theory (EMT) [20,21] the authors suggested that the rise in capacitance ($C$) can be explained as a result of increasing effective surface areas ($A$) of metallic clusters and decreasing the distances ($d$) between them (since $C \sim A/d$). However, in our samples, comparison between the magnitudes of dielectric response (Figs. 1a, 1b) (and the magnitude itself) suggests that this effect alone can not explain the several order rise in dielectric permittivity. The EMT is valid when fluctuations in the local microscopic values of dielectric permittivity are small, which is not the case in our compounds where resistivity of two phases differs by a huge factor. We, therefore apply generalized Maxwell-Garnett theory (MGT) [22] for macroscopic analysis. At low temperature in a zero magnetic field the 37% doped sample is an antiferromagnetic insulator. When magnetic field is applied the ferromagnetic conductive phases start to appear in antiferromagnetic insulating matrix [23]. The expression for effective dielectric permittivity $\varepsilon_{eff}$ can thus be given by [22]:

$$\frac{\varepsilon_{eff}(\omega) - \varepsilon_i(\omega)}{F\varepsilon_{eff}(\omega) + (1 - F)\varepsilon_i(\omega)} = \nu \frac{\varepsilon_m(\omega) - \varepsilon_i(\omega)}{F\varepsilon_m(\omega) + (1 - F)\varepsilon_i(\omega)}, \quad (1)$$

where $F$ is the depolarization factor which for spherical grains is $1/3$. The spherical shape of grains was recently confirmed by small angle neutron scattering study [24]. $\varepsilon_i$ dielectric permittivity of antiferromagnetic phase, $\varepsilon_m$ dielectric permittivity of ferromagnetic fraction, $\nu$ is the volume fraction of ferromagnetic phase, which can be determined from magnetization loop [23]. When magnetic field is continuously increasing, $\nu$ rises also and finely reaches 100%. At a certain point it is no longer valid to consider a model in which ferromagnetic metallic phases are surrounded by an antiferromagnetic insulating matrix. In fact it is rather the opposite: smaller antiferromagnetic insulating phases are embedded in a ferromagnetic conductive environment. This situation occurs at $\nu \cong 0.4$ [22]. As a consequence when the system approaches to the metal-rich limit it is obvious that $\varepsilon_i$ and $\varepsilon_m$ should be interchanged in equation (1). Consequently, magnetic moment jumps also with a little delay at IM transition (Fig. 3). The bigger difference in dielectric permittivity between insulating and conductive phases the larger change in effective dielectric permittivity is expected. In addition, an anomaly appears when the system reaches the metal (insulator) rich limit (Figs. 1, 2b, 2c). In the case of the $Pr_{0.7}Ca_{0.3}MnO_3$ crystal the ground state at zero field consist of ferromagnetic insulating phases in antiferromagnetic insulating matrix at low temperatures [24]. It is also quite reasonable that for the 30% doped sample the rise in dielectric permittivity is smaller, since less energy is required (lower magnetic field) to induce IM transition (Fig. 2). This is also probably why the jump in resistivity of the sample is also smaller (Fig. 2) [12]. Qualitative analysis of equation (1) indeed predicts increasing of effective dielectric permittivity as a function of volume fraction in a phase separated samples. However, quantitative result differs radically with experimental data, strongly suggesting that additional mechanisms are involved as well. For example, taking $\varepsilon_i = 35$, $\varepsilon_m = 55000$, $\nu = 0.11$ the resulting dielectric permittivity at 4.1 T will be 48, while experiment gives around 20 000 (Fig. 3). It also has to be mentioned that small modification of the metallic phase under applied magnetic field was recently suggested by magnetic field dependence of polaron activation energy [25]. Although MGT theory predicts additional contribution to the effective dielectric permittivity as a volume of two fractions changes, it still can not explain a huge value of dielectric permittivity of the metallic phase in both compounds. Therefore, microscopic contribution to the effective dielectric permittivity has to be also taken into account. The theoretical study on MI transition is still a subject of many works [26–29]. Among various proposed mechanisms, the role of electron-electron interactions in the formation of insulating (or conducting) phases has received a great deal of attention [30–33]. It was pointed out by Hubbard [34] that MI transition happens when the interaction energy between electrons reaches some critical value $U_c$ (Mott-Hubbard transition). In the case of our compounds one can assume existence of vice versa mechanism leading to the screening of Coulomb interaction under magnetic field at IM transition. Thus, our crystals can be considered as so called Mott-Hubbard insulators at low temperature ground state. The effects of polarization and screening become more and more pertinent as the IM transition is approaching. This situation may be analysed by adding to Mott-Hubbard Hamiltonian an additional term describing coupling to the static electric field [35]:

$$H = -t \sum_{<i,j>\sigma} C_{i\sigma}^{\dagger} C_{j\sigma} + U \sum_i \widehat{n}_{i\uparrow} \widehat{n}_{i\downarrow} - E \sum_i x_i \widehat{n}_i, \quad (2)$$

where the first term describes electron hopping, $t$ is the nearest-neighbour site hopping integral. The $C_{i\sigma}^{\dagger}(C_{i\sigma})$ are Fermi-operators of creation (annihilation) of the electron at $i$th site with a spin projection ($\sigma n_{i\sigma} = C_{i\sigma}^{\dagger} C_{i\sigma}$ operator describing quantity of the electrons at the site with a given spin projection up↑ and down↓). Second term represents the electron correlation energy contribution ($U$ is the Coulomb repulsion energy). In the third term $x_i$ is the coordinate of the $i$th site, $\widehat{n}_i$ is the $i$th site occupation, $x_i\widehat{n}_i = \hat{X}$ is the dipole operator. It has to be pointed out that if the value $2tz$ (where $z$ is the coordination number) is much bigger than $U$, the system is a Fermi-liquid. If $2tz \ll U$ system is strongly correlated. Using perturbation theory, electric susceptibility under open boundary conditions can be related to the states $|\psi_n\rangle$ of the Hamiltonian (2) by [35]:

$$\chi = 2L^{-d} \sum_{n \neq 0} \frac{|\langle\psi_0|\hat{X}|\psi_n\rangle|^2}{\delta E_n}, \tag{3}$$

$$\chi \leq \frac{2}{\Delta} L^{-d} \langle\psi_0|X^2|\psi_0\rangle, \tag{4}$$

where, $\Delta$ is the minimum excitation energy for which the dipole matrix element still exists (the so called charge gap), $d$ is the dimension of the lattice with linear dimension $L$ and

$$\langle\psi_0|X^2|\psi_0\rangle = \sum_S \langle\psi_0^{(S)}|X^2|\psi_0^{(S)}\rangle \approx \langle S\rangle l^2, \tag{5}$$

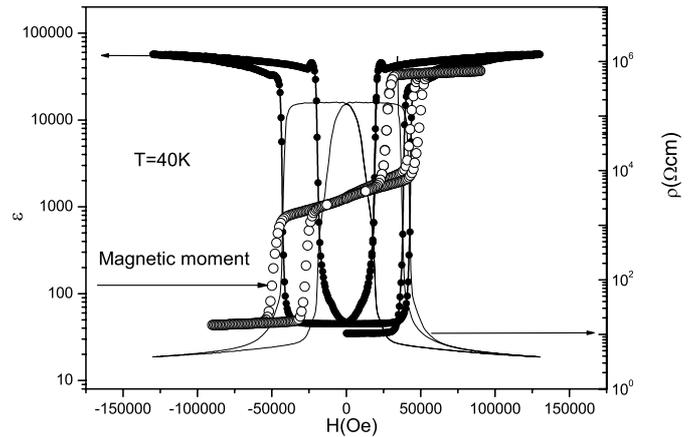

**Fig. 3.** Dielectric permittivity (plot on the left scale), resistivity (plot on the right scale) and magnetization versus magnetic field for $Pr_{0.63}Ca_{0.37}MnO_3$ crystal.

$S$ is the number of doubly occupied sites. At high Coulomb repulsion energy (insulator state), $e_g$ electrons are localized close to $Mn^{3+}$ ions (i.e., hopping is minimized) and formation of randomly oriented dipoles can occur of the average size $l$. Comparing (4) and (5) one can judge that apart from microscopic effects the huge increase of dielectric permittivity in our samples can be due to either the increase of the size of inequivalent Mn dipoles (and off-centering [36]) or a decrease in the charge gap. The scenario when all aforementioned mechanisms contribute can not be excluded when taking into account the enormous magnitude of the effect.

It is worth mentioning that in both samples the slope of dielectric permittivity changes with temperature in a comparable way, following the magnetic phase diagram of each compound. Magnetic-field-induced relative change of dielectric permittivity decreases at higher temperatures (Fig. 4). In the case of $Pr_{0.7}Ca_{0.3}MnO_3$ sample, dielectric permittivity reveals remanence in a low temperature region (<60 K) (Fig. 2), where clear magnetic contribution of Pr moments was previously observed by neutron scattering studies [37]. Further substitution of Pr ions with Ca (in the higher doped $Pr_{0.63}Ca_{0.37}MnO_3$) naturally leads to the disappearance of this effect at lower temperatures (Figs. 1b and 5).

## 4 Conclusions

In summary, a several order increase of dielectric permittivity was found in both samples of $Pr_{1-x}Ca_xMnO_3$ crystals with different doping levels ($x = 0.30, 0.37$). This is an evidence for dielectric catastrophe phenomenon naturally taking place at the IM transition. The same effect is expected to be seen in the system at IMT induced by pressure, light and X-ray. In the case of 0.37 doped sample the effect is bigger since more energy is needed to overcome critical Coulomb repulsion energy and to get created microscopic metallic phases, which then additionally contribute to the magnetodielectric effect through MGT theory. Macroscopic MGT and extended Mott-Hubbard theory jointly are proposed to describe the experimental results. In light of unique resistivity correlated behaviour of dielectric permittivity near IM transition it may be interesting to see how much dielectric properties of other manganites (for example [38–41]) are correlated with the colossal magnetoresistance effects in these materials.

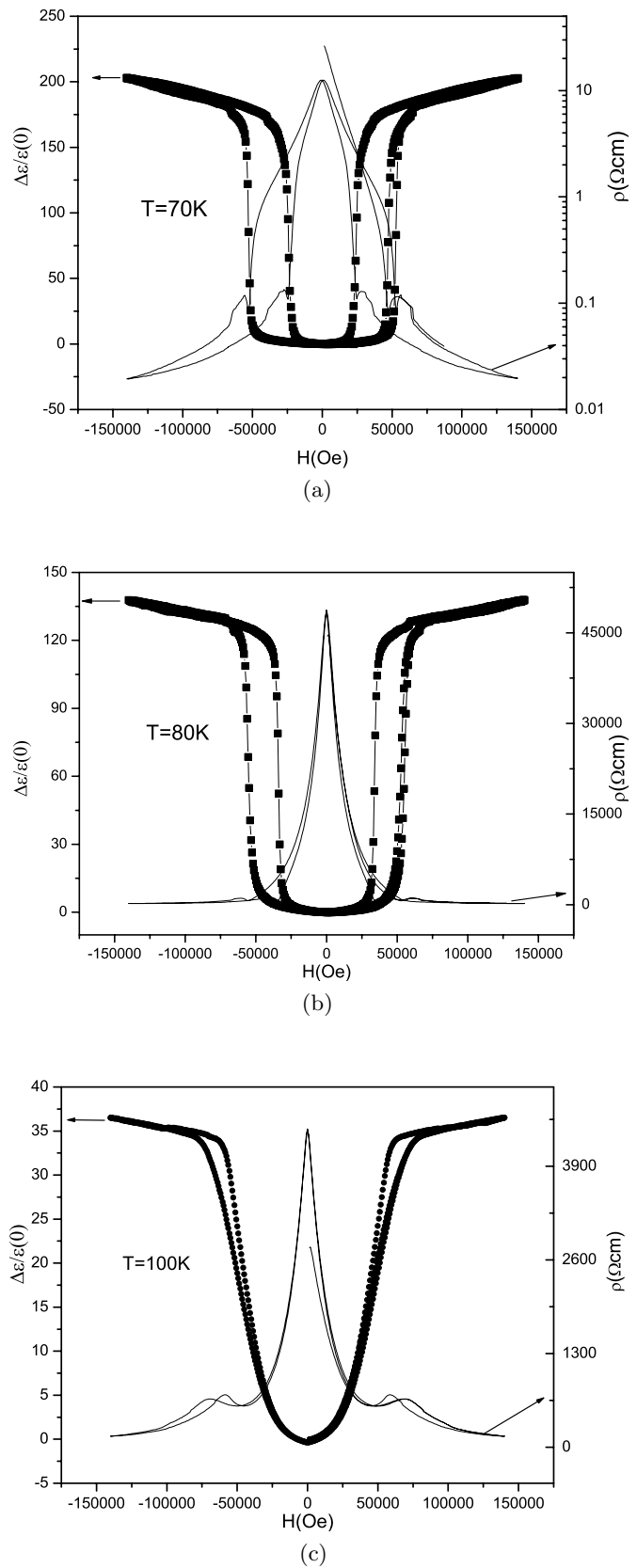

**Fig. 4.** Magnetodielectric and resistivity isotherms $Pr_{0.7}Ca_{0.3}MnO_3$ crystal taken at 70, 80 and 100 K. Symbolic and solid loops refer to the left and right scale respectively.

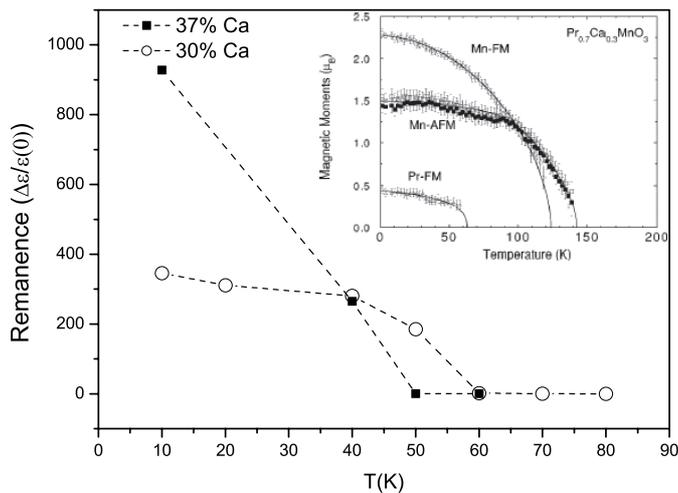

**Fig. 5.** Dielectric remanence as a function of temperature. Inset: magnetic moments as a function of temperature for $Pr_{0.7}Ca_{0.3}MnO_3$ reproduced from reference [37].

We thank L. Hervé for crystal growth. Helpful discussions with Dr. H. Eng, Dr. D. Sinclair and Dr. M.P. Singh are gratefully acknowledged. We also thank NATO science programme and FAME NOE (FP6-500159-1) programme for financial support.